\documentstyle[12pt]{article}
\newcommand{\beq}{\begin{equation}}
\newcommand{\eeq}{\end{equation}}

\def\op{operator}

\def\cn{condition}

\begin{document}
\begin{titlepage}
\begin{flushright}
NBI-HE-96-45\\
HUTP-96/A038\\
August 1996\\
\end{flushright}
\vspace{0.5cm}
\begin{center}
{\large {\bf A ${\bf Z}_2$ Classification for\\
2D Fermion Level Crossing}}\\
\vspace*{1.2cm}
{\bf Minos Axenides}\footnote{e-mail: {\tt axenides@talos.cc.uch.gr}}\\
\vspace{0.2cm}
{\em University of Crete, Department of Physics\\
 GR-71409 Iraklion-Crete, Greece}\\
\vspace*{0.4cm}
{\bf Andrei Johansen}\footnote{e-mail: {\tt johansen@string.harvard.edu},
{\tt ajohansen@nbivax.nbi.dk}}\\
\vspace*{0.2cm}
{\em Harvard University, Lyman Laboratory of Physics\\
Cambridge, MA 02138, USA}\\
\vspace*{0.4cm}
{\bf Holger Bech Nielsen}\footnote{e-mail: {\tt hbech@nbivax.nbi.dk}}\\
\vspace*{0.2cm}
{\em The Niels Bohr Institute, 
University of Copenhagen\\ Blegdamsvej 17, DK-2100 Copenhagen \O, Denmark}
\end{center}
\vspace{1.5cm}

\begin{abstract}

We demonstrate that the number of fermionic zero modes of the static 
$2$-dimensional Dirac operator in the background of $SU(2)$ static
gauge-Higgs field configurations is a topological invariant modulo four.
Static configurations which are everywhere odd under parity
with even-parity pure gauge behaviour at infinity
admit $4n$, $n\in {\bf Z},$ zero modes of the 
Jackiw-Rebbi (JR) type. Odd-parity configurations with odd-parity pure
gauge behaviour at infinity are
topologically disconnected from the vacuum and admit 
$4 n + 2$ 
fermionic zero energy solutions. The classification implies the 
collapse of half of the
fermion zero modes upon embedding a 
$2$-dimensional
gauge-Higgs configuration (string) with odd-parity pure
gauge behaviour at infinity into the $3$-dimensional 
Minkowski space.

\end{abstract}
\end{titlepage}
%\newpage
\section{Introduction}

Anomalous fermion level crossing in gauge theories is intrinsically connected
to their topologically nontrivial periodic vacuum structure. For theories
with the gauge symmetry spontaneously broken there appears a potential barrier
between the topologically inequivalent vacua which is proportional to the
mass scale of the theory. At finite temperatures thermally induced transitions
above the barrier are accompanied by fermion level crossing. It is due to the
presence of static gauge-higgs fields with Chern-Simons number $CS=1/2$ , the
sphalerons, which are saddle points to the field equations. In the context of
the standard $SU(2)\times U(1) $ electroweak theory such configurations induce
an odd number of normalizable zero energy solutions to the 3D Dirac equation.

It was previously found \cite{MA,JM} that for the gauge
fields $ A_{i}$ having 
odd parity everywhere the Chern-Simons number is a topological charge. 
In a gauge
where $A_{i}$ is odd under parity transformation the $U$-odd configurations
(where $U$ is defined via the asymptotic behaviour of $A_{i}$ at infinity,
$A_{i}\to -i\partial_{i}U(x)U(x)^{-1}$) have $CS= n+1/2$,
$n\in {\bf Z}$, and admit an odd number of fermionic zero modes. 
They contribute to unsuppressed $B+L$ violating thermal transitions. 
The $U$-even configurations have
$CS=n$ with an even number of fermionic zero modes. 
They are 
homotopically equivalent to
the vacuum and are not directly connected to the $U$-odd
configurations.
In other words the $3D$ Dirac operator in the background of odd-parity 
gauge-Higgs fields admits a
topological invariant number 
modulo 2 of zero modes.
The presence of 
$U$-odd configurations is a consequence of a nontrivial 
$Z_{2}$ structure in the
space of static finite energy configurations in the electroweak theory.
It follows from the existence of the homotopy groups $[S^{2}/{\bf Z}_2 , 
SU(2)/{\bf Z}_{2}] = {\bf Z}_{2}$ and $[S^{3}/{\bf Z}_{2}, SU(2)/{\bf Z}_{2}]=
{\bf Z} \times {\bf Z}_{2}$ \cite{JM}.

In the present note we examine the eigenvalue spectrum of the static two dimensional  Dirac
operator in the background of $SU(2)$ and $U(1)$ gauge fields respectively. 
The presence
of zero modes in the background of embedded string loops of the Nielsen-Olesen
type \cite{NO} was first demonstrated by Jackiw and Rebbi \cite{JR}. 
For 
$Z$-string loops the issue was studied by Earnshaw and 
Perkins \cite{EP}.

The organization of the paper is the following.
In section 2 we discuss the $SU(2)$ theory in (2+1) dimensions and 
the problem of extension of its topological defects into (3+1) 
dimensions.
In section 3 we discuss the $U(1)$ gauge theories in  different 
dimensions and the problem of embedding the low dimensional topological defects into higher dimensions.
We close by summarizing the results of the paper.

\section{SU(2) gauge theory in (2+1) dimensions}

We discuss the spectrum of the Dirac operator in 
the background of a static configuration.

To that end we consider the 2D Dirac equation for 
the Dirac 2D fermion
in an external gauge field $A_i(x)$
\beq
\sigma_i \nabla_i \psi =
\sigma_i (\partial_i -i A_i (x)) \psi (x) = i\lambda \psi (x) ,
\eeq
where $i=1,2 .$
The matrix $\sigma_3$ plays the role of the
4D $\gamma_5$ 
since it anticommutes with all 2D $\sigma$ matrices.
Thus the Dirac spinor can be split into two one component chiral spinors
$\psi = (\psi_1 ,\psi_2) .$
It is worth emphasizing that a chirality cannot be defined in 3D.
The anticommutation property of the $\sigma_3$ matrix  
automatically means that in 2D for any spinor $\psi$ which obeys 
eq.(1) one can construct the spinor $i\sigma_3 \psi$ which
corresponds to the eigenvalue $-i\lambda .$
We thus prove that in 2D all non-zero eigenvalues are paired
$(i\lambda , -i\lambda) .$
Moreover for the case of 
an $SU(2)$ gauge group we can choose  
the eigenfunction for any eigenvalue $\lambda$ to be ``real''
\beq
i\sigma_2 i\tau_2 \psi^* (x) = \psi (x),
\eeq
while the real eigenfunction for $-i\lambda$ reads $i\sigma_3 \psi (x) .$
Here $\tau_i$, $i=1,2,3$ are the isospin
matrices.

It is easy to see that there is also
a doubling of zero modes of the Dirac operator.
Indeed let us consider $(\psi_1 ,0)$ and $(0,\psi_2)$ as a basis of the 
kernel of the Dirac operator.
For these components we have the following equations
\beq
\nabla \psi_2 = 0,\;\;\; \bar{\nabla} \psi_1 = 0.
\eeq
Here $\nabla = \nabla_1 -i\nabla_2$ and 
$\bar{\nabla} = \nabla_1 + i\nabla_2 .$
It is now easy to see that for any $\psi_1\neq 0$, $\bar{\nabla} 
\psi_1= 0$, one can explicitely 
construct $\psi_2=i\tau_2\bar{\psi}_1\neq 0$, $\nabla \psi_2 = 0$,
where $\tau_2$ is the isospin
matrix.
This proves the doubling of zero modes.

In the case of an odd-parity 
gauge field, $A(x)=-A(-x)$, 
we have an additional degeneracy for all the non-zero eigenvalues.
Let $\psi$ be a real eigenfunction for an non-zero eigenvalue 
$i\lambda .$
We can easily check that the real spinor $i\sigma_3 \psi (-x)$ corresponds
to the eigenvalue $i\lambda$ while $\psi (-x)$ corresponds to 
$-i\lambda .$
Moreover the eigenfunctions constructed above are linearly independent.
Indeed let us assume that the wave functions $\psi$ and 
$i\sigma_3 \psi (-x)$ were linearly dependent, i.e. 
$\psi (x) = a\sigma_3 \psi (-x)$, where $a$ is a constant.
Then it is easy to see $a=\pm 1$.
Let us now consider the reality condition of $\psi$, eq.(2).
By combining it with the condition of linear dependence 
one can get that $\psi (x) = \pm i\sigma_3 \psi (-x)$. 
It clearly
contradicts with $a=\pm 1$, if $\psi\neq 0$.
We thus have that each eigenfunction with non-zero $\lambda$ generates
a different eigenfunction for the same $i\lambda .$
Hence each non-zero eigenvalue is twice degenerate, and moreover there is 
a pairing of levels $(i\lambda , -i\lambda ).$
This doubling for non-zero modes is of course a result of
the odd parity behaviour of the gauge field.

Let us consider the zero modes of the above Dirac \op \
in an odd-parity gauge field.
If there is a zero mode 
we can then construct the above 
four wavefunctions 
which obey the same equation.
However such a wave function can be odd or even with respect to 
a parity transformation $\psi (-x) =\pm \psi (x) .$
Therefore the number of linearly independent zero modes may be less 
than 4.
However it is easy to see that the number of linearly independent 
zero modes is always even. 
Indeed for any real zero mode
with a wave function $\psi (x)$ we can separately 
construct a linearly independent one 
$i\sigma_3 \psi (x)$ (as discussed above).

Thus we see that the number of zero modes of the Dirac operator
is invariant modulo 4 (but not modulo 2 as we 
had in 3 dimensions) when we smoothly change 
the external odd-parity gauge field.
Indeed if we change smoothly the external (odd under parity) gauge field
a non-zero eigenvalue $i\lambda$ can cross zero.
At some moment $i\lambda$ gets to be zero while by continuity the four above
eigenfunctions being linearly independent (the wave functions for 
different values of $\lambda$ are orthogonal) are zero mode wave functions.
Hence under smooth deformations of the gauge field (odd under parity) 
the number of zero modes can change only by $4k ,$ where $k$ is an integer. 

It is worth emphasizing that 
the quadroupling of the 
non-zero modes is of course a result of the
odd parity behaviour of the gauge field as well as 
of the above index theorem 
modulo 4.
In the case of non-constrained gauge fields (without a definite
parity) there is no index theorem since the number of zero modes is always 
even while the non-zero modes are paired and there is no 
additional doubling for them.
This corresponds to the trivilatity of the homotopy group
$\pi_1(SU(2))=0$.

Thus at least for the gauge group $SU(2)$ the topological classes
of parity odd gauge fields are two classes which allow for
the number of fermionic zero modes to be 
2 mod 4 and 0 mod 4, respectively.
Following the approach of ref. \cite{MA,JM} one can 
now show that these
classes correspond to the odd-parity gauge fields with 
even and odd pure gauge behaviour at ininity, $A_i(x)\to 
-i(\partial_i U U^{-1})(x)$, $U(x)=\pm U(-x)$.
Since a vacuum configurationt obviously does not allow for fermionic zero 
modes and has $U(x)=U(-x)$ behaviour at infinity we conclude that 
the sector of gauge fields with such a behaviour requires 0 (modulo 4)
fermionic zero modes.
Thus the homotopically non-trivial sector corresponds to
gauge fields with $U(x)=- U(-x)$.
These fields require 2 (modulo 4) fermionic zero modes.

The above group element $U(x)$ is a map from $S^1$ into $SU(2)$. 
Thus we have proved that there is a non-trivial homotopy group \cite{textbook}
\beq
[S^1/{\bf Z}_2 , SU(2)/{\bf Z}_2] = {\bf Z}_2 .
\eeq
This group is responsible for the above index theorem.
Formally eq. (4) follows from the facts that $S^1/{\bf Z}_2$ is 
isomorphic to $S^1$ and $SU(2)/{\bf Z}_2 = SO(3) .$

We will
now discuss an embedding of two-dimensional 
topologically 
non-trivial objects, such as strings,
into three dimensional space.
In particular we consider the 2D Dirac operator in the external field of the
Nielsen-Olesen string, i.e. we consider a 2D slice  
perpendicular to the line of the vortex. 
The fermionic zero mode in such a field 
is proportional to $1/r$ at infinity ($r$ is the 
distance to the line of the vortex).
Hence the normalization integral is logarithmically divergent.
Nevertheless 
these zero modes are still relevant if we assume that the 2D space
is compactified. 
For example the simplest way to compactify the space
is to introduce
an external metric which makes the normalization integral convergent.
In this case we have a couple of zero modes \cite{JR}
that agrees with the above analysis.

We now construct a loop of such a string.
Since any configuration in three dimensional space is contractable
such a loop may allow for any number of fermionic zero modes.
Therefore we have to formulate the prescription for
making a
three dimensional object if we want to find a correspondence between
the homotopy properties of two and three dimensional configuration.
We shall assume that the radius of the loop is much larger than the size of 
its core.
We will also assume that we construct a three dimensional gauge field
configuration which is odd under parity transformation.
In this case there are two different situations as it has been 
demonstrated in ref. \cite{MA,JM}.
One of two sets of three dimensional 
configurations allows for an even number of fermionic zero modes.
In this case one considers a twisting \cite{vach} of the string
which corresponds to pure gauge behaviour of the 
three dimensional gauge field $A_i (x)\to -i(\partial_i U U^{-1})(x)$ 
at infinity with even $U(x)=U(-x).$
The second set of three dimensional fields corresponds to 
a non-trivial twisting of the string. 
The resulting 
three dimensional gauge fields have pure gauge behaviour at infinity
with odd $U(x)=-U(-x)$.
These fields allow for an odd number of fermionic zero modes.

From the analysis of \cite{MA,JM} it follows that 
some fermionic zero modes which existed on the 
two-dimensional slice do not cause an appearance of 
zero modes of the three dimensional Dirac operator in the string loop
configuration.
Therefore we conjecture that a non-trivial twisting of the string
configuration (odd-parity pure gauge behaviour at infinity)
kills half of the zero modes of the Dirac operator 
since 
a three dimensional Dirac operator has an odd number of
normalizable zero modes in this case.
The reason for the disappearance of half of the zero modes is probably 
that only one of their linear combinations can satisfy the 
condition of single-valuedness of the wavefunction
when we extend the two dimensional object into
three dimensions.
In the opposite case of a topologically trivial twist 
(even-parity pure gauge behaviour at infinity) all zero modes
might survive.

We thus conjecture that there is a 
correspondence of the ${\bf Z}_2$ structure 
of the theory
in (2+1) dimensions to that of the (3+1) dimensional theory.
This correspondence 
is non-trivial because it depends on the twisting.
For an appropriate twisting the non-trivial homotopy class in (2+1) dimensions 
maps to the non-trivial homotopy class in (3+1).

\section{U(1) gauge theory in (3+1), (2+1) and (1+1) dimensions}

Here we shall consider the gauge theories with $U(1)$ gauge group.

Let us first start with 
a gauge theory in (1+1) dimensions defined on
$I\times {\bf R},$ where the interval $I=[-\pi,\pi]$ stands for 
the ``space''.
In this case the homotopic properties of static configurations are 
described by
the fact that the homotopy group of maps $[\partial I,U(1)]=0$,
where $\partial I$ consists of the two end points of the interval
$I$.
Therefore there is no index theorem for the static Dirac operator
in this case.
Let us now turn to the static gauge fields even under the parity 
transformation.
This restriction is due to the fact that the sphaleron 
configuration in (1+1) dimensions is a constant gauge field (see e.g. 
\cite{shaposhnikov}).
In this case we have the following non-trivial homotopy group
$[\partial I/{\bf Z}_2 ,U(1)/{\bf Z}_2]={\bf Z}_2 .$ 
Therefore there are two classes of gauge fields which are not 
continuously related to each other: with odd-parity and even-parity pure 
gauge behaviour at the ends of the space interval $I=[-\pi,\pi]$.
A particular representative of the non-trivial homotopy class is just 
the above sphaleron $A_x=1/2$.
More generally $A_x=n/2$, where $n$ is an odd-integer.
For the trivial class we have $n$ to be even. 
The existence of a non-trivial homotopy group 
$[\partial I/{\bf Z}_2 ,U(1)/{\bf Z}_2]={\bf Z}_2$ implies the existence
of an index theorem for the Dirac operator in 1 dimension in the 
presence of even-parity gauge fields.
The Dirac operator has one zero mode for odd $n$ and 
0 zero modes for even $n$ (we assume the antiperiodicity of the 
fermionic wavefunctions).

For the $U(1)$ gauge theory in (2+1) the relevant homotopic group formally reads
\beq
[S^1/{\bf Z}_2 ,U(1)/{\bf Z}_2] ={\bf Z} 
\eeq
since $U(1)/{\bf Z}_2 = {\rm R}P^1 =U(1) .$
If we denote an element of 
the $U(1)$ group as $\exp i\phi (\theta) ,$ where
$\theta$ is a coordinate on $S^1 ,$ $\theta \in [0, 2\pi] ,$ then the 
classes of the above homotopic group are represented by group elements
for which $\exp 2i\phi (\theta)$ is periodic on $[0,\pi].$
This group element can have any integer winding number on the interval
$[0,\pi ] .$
This is the same winding number as for the element $\exp i\phi (\theta)$
on the interval $[0,2\pi ].$
Therefore all
the classes of ${\bf Z}$ are split by parity into two subgroups: with odd
and even winding numbers, respectively.

Let us now consider the spectrum of the Dirac operator. 
In the case of a gauge group $U(1)$ the eigenfunctions 
cannot be generally chosen real.
For a generic gauge field there is no doubling of zero modes and
the number of zero modes can be both odd and even in contrast to the above 
example.
However there is a pairing of non-zero wave functions given by 
$\psi (x)$ and $i\sigma_3 \psi (x)$ for eigenvalues $\lambda$ and $-\lambda$
respectively.
We see that the number of zero modes can change only by an even number
under smooth deformations of gauge fields due to a pairing of non-zero levels.
Hence we have an index theorem that the number of fermionic zero modes
is invariant modulo 2.
Actually in this case it is well known 
that the index of the 2D Dirac operator is
given by the winding number of the gauge field \cite{textbook}.

Let us now 
consider the case of an odd under parity 
gauge field.
Because of non-reality the wave functions $\psi (x)$ and 
$i\sigma_3 \psi (-x) $ which correspond to the same eigenvalue $\lambda$
can be linearly dependent.
Therefore in general there is no doubling of non-zero modes.
The same is obviously true for zero modes.
Indeed, given a wave function $\psi (x)$ for a zero mode we
can construct three more wave functions $\psi (-x) ,$ 
$i\sigma_3 \psi (x)$ and $i\sigma_3 \psi (-x) .$ 
It is easy to see that the \cn \ of 
linear dependence of all these 4 wave functions is not contradictory.
Therefore we see that when we constrain the gauge fields to be odd under parity
the index theorem mod 2 for zero modes still holds without any changes.
This fact agrees with the above statement that no homotopy classes appear 
due to oddness under parity.
Instead we get a classification of 
the usual ${\bf Z}$ classes: with odd and even 
winding numbers.
The sense of the index theorem can be expressed as follows: it is not possible
to change the number of zero modes from odd to even and back under a smooth
deformation of the gauge field.

We conclude that the number of zero modes of the Dirac operator 
is invariant modulo 2.
Thus it is odd in the non-trivial class of the gauge fields
and even in the trivial class with respect to the above splitting of
${\bf Z}$ into $2{\bf Z}$ and $2{\bf Z}+1 .$
Such a conclusion agrees with the index theorem for the Dirac \op \ which is 
coupled to a $U(1)$ gauge field, which states that the difference of 
numbers of left-handed and right-handed zero modes is given by a winding 
number of the gauge field. 

Let us now consider an embedding of (1+1) static configurations into 
(2+1) dimensions.  
Under such an embedding the coordinate on 
an interval $I$ is naturally identified with $S^1$ which corresponds to 
the polar angle $\theta$ in two dimensional space. 
Then it is easy to see that odd fields $A_x=(n+1/2)$ cannot be
embedded since they would correspond to a not single-valued 
group element at infinity.
Hence only gauge fields $A_x=n$ can be embedded.
Thus the spectrum of the Dirac operator reduces under such an embedding
since half of the zero modes of the 1-dimensional Dirac operator become 
non-single-valued. 
The above ${\bf Z}\to 2{\bf Z}\oplus 2{\bf Z}+1$
homotopic classification in (2+1) dimensions corresponds to the odd-parity
gauge fields with even or odd pure gauge behaviour at infinity.
Thus there is no connection between the ${\bf Z}_2$ structure in (1+1) and
(2+1) dimensions under such an embedding.

We now turn to the 3D $U(1)$ gauge theory.
Since the group $\pi_2(U(1))=0$ there is no index theorem for the Dirac 
operator.
We firstly consider 
the sector of odd-parity gauge fields.
The relevant homotopy group for this case is also trivial $[{\rm R}P^2 , 
U(1)]= 0.$
This follows from the fact that \cite{textbook}
\beq
[RP^2 ,U(1)] = {\rm Hom} ({\bf Z}_2 ,{\bf Z}) =0.
\eeq
We also present here a simple intuitive explanation of this result.
The ${\rm R}P^2$ sphere is equivalent to a hemisphere with identified 
opposite points on the boundary.
Formally by
following what we did 
in the $SU(2)$ case we get two classes of gauge fields
which are odd under parity: the ones
with odd and even respectively pure
gauge behaviour at infinity, i.e. on $S^2 .$
These $U(1)$ group elements are given by exponentials of imaginary 
functions $\exp i\alpha (x) .$
We want now to show that the class of odd $g =\exp i\alpha (x)$ is empty.
Let us consider the boundary of the above hemisphere: $g$ is odd on this 
boundary which is $S^1 .$
Then by continuity it is clear that $\alpha (x)$ is shifted by $2\pi$
when moving around this boundary.
Due to 
the non-triviality of $\pi_1 (U(1)) ={\bf Z}$ this group element 
which of course belongs to a non-trivial class of $\pi_1 (U(1))$
is not contractable to unity.
Hence when extended to the whole 
hemisphere there appears a singularity
at some point on it. 
Therefore we conclude that the class of odd $g$ is empty
because by definition we have to consider smooth group elements on 
$S^2 .$
Such an analysis could not prove the absence of odd classes for the
$SU(2)$ group because $\pi_1 (SU(2)) =0 .$

Thus there is no non-trivial classes in 
the 3D case for the group $U(1) .$

Let us now consider the case of 
an $SU(2)\times U(1)$ gauge group
in (2+1) and (3+1) dimensions.
In (2+1) dimensions there is a non-trivial ${\bf Z}_2$ structure 
which depends on the embedding of the
${\bf Z}_2$ group into $SU(2)\times U(1)$.
In particular there is a non-trivial ${\bf Z}_2$ structure if ${\bf Z}_2
\in U(1)$.
However as it was shown in ref. \cite{JM} the corresponding homotopy
group $[{\rm R}P^2,SU(2)\times U(1)/{\bf Z}_2]=0$ for the case ${\bf 
Z}_2\in U(1)$.
Therefore this part of the ${\bf Z}_2$ structure of 
a (2+1) theory is lost under
the embedding into (3+1) theory.
Therefore only the ${\bf Z}_2$ structure that 
corresponds to the center of $SU(2)$  
survives under this embedding.
 
\section{Conlusions}

We extended our previous study of homotopic 
classification of odd-parity deformed sphalerons to lower dimensions.
Remarkably enough we found for the $SU(2)$ gauge theory 
a non-trivial ${\bf Z}_2$ structure in (2+1) dimensions which 
implies an index theorem modulo 4 for the Dirac operator.
This implies 
that the number of zero modes of the Dirac operator
is a topological invariant which takes values (0 or 2) modulo 4.
This corresponds to the existence of the non-trivial homotopy group 
$[S^1/{\bf Z}_2,SO(3)]={\bf Z}_2 .$

We argued for the existence of a 
non-trivial correspondence 
of such a ${\bf Z}_2$ structure for static
gauge fields 
in two and three dimensions.
More specifically
the embedding of a 2-dimensional 
static configuration, such as a Nielsen-Olesen 
vortex, into a 3-dimensional one, such as 
a string loop results in the collapse of an odd number of zero modes
from the 2 mod 4 it may otherwise possess.

We also investigated the ${\bf Z}_2$ structure in the
$U(1)$ gauge theories in (1+1), (2+1) and (3+1) dimensions.
We argued that there is no non-trivial map of the ${\bf Z}_2$ structure
from the (2+1) dimensional theory into the (3+1) dimensional one, 
and also 
there is no non-trivial map of the ${\bf Z}_2$ structure
from the (1+1) dimensional theory into the (2+1) dimensional one.

For its potential physical 
implications the first part of our results is the more 
relevant. 
It suggests 
that the existence of zero modes of the Dirac 
operator in 2 dimensions does not imply the existence of zero 
modes in three dimensions. 
Thus we see that the B-violating properties of three 
dimensional configurations,
such as the electroweak strings, are not directly 
related to the homotopic properties of 
the two dimensional configurations
which are embedded into the three dimensions. 

\section{Acknowledgments}
A.J. is grateful to the Niels Bohr Institute for warm hospitality.
His research is partially supported by a NATO grant GRG 930395.
M.A. 
acknowledges support from the European Union 
contract number ERBCHRXCT-940621.


\begin{thebibliography}{8}
\bibitem{MA} M. Axenides and A. Johansen, Mod. Phys. Lett. {\bf A9}
(1994) 1033; 
M. Axenides, A. Johansen, H. B. Nielsen and O. Tornkvist, Nucl. Phys. {\bf B 474} (1996) 3.
\bibitem{JM} M. Axenides, A. Johansen and J. Moller,  
J. of Math. Phys. {\bf 36} (1995) 5284.
\bibitem{NO}H.B. Nielsen and P. Olesen, Nucl. Phys. {\bf B 61} (1973) 45.
\bibitem{JR} R. Jackiw and C.Rebbi, Phys. Rev. {\bf D 13} (1976) 3398.
\bibitem{EP} M. Earnshaw and W. Perkins, Phys. Lett. {\bf B 328} (1994) 337.
\bibitem{textbook} See e.g. B.A. Dubrovin, A.T. Fomenko and S.P. Novikov,
{\it Modern Geometry - Methods and Applications, Part II: The Geometry
and Topology of Manifolds} (Springer-Verlag 1985).
\bibitem{vach}T. Vachaspati, Phys. Rev. Lett. {\bf 68} (1992) 1977;
M.Barriola, T. Vachaspati and M. Bucher, Phys. Rev. {\bf D 50} (1994) 2819.
\bibitem{shaposhnikov}V.A. Rubakov and M. E. Shaposhnikov, hep-ph/9603208, to appear in Usp. Fiz. Nauk
{\bf 166} (1996).

\end{thebibliography}
\end{document}